


\input harvmac
\input epsf

\def\ga{\gamma}

\def \la{\lambda}

\def \th{\theta}

\def\taub{{\overline{\tau}}}

\def\tautilde{\tilde\tau}

\def\nutilde{\tilde\nu}
\def\gatilde{\tilde\ga}

\def\btilde{\tilde b}
\def\dtilde{\tilde d}
\def\etilde{\tilde e}
\def\ttilde{\tilde t}
\def\utilde{\tilde u}

\def\bbar{{\overline{b}}}

\def\tbar{{\overline{t}}}

\def\5bar{{\overline{5}}}

\def\sy{supersymmetry}
\def\sic{supersymmetric}

\def\ssm{supersymmetric standard model}

\def\pa{\partial}
\def\321{$SU_3\otimes\ SU_2\otimes\ U_1$}

\def\ijmpa{{Int.\ J.\ Mod.\ Phys.\ }{\bf A}}

\def\npb{{Nucl.\ Phys.\ }{\bf B}}

\def\plb{{Phys.\ Lett.\ }{\bf B}}

\def\prd{{Phys.\ Rev.\ }{\bf D}}
\def\prl{Phys.\ Rev.\ Lett.\ }

\def \in{\leftskip = 40 pt\rightskip = 40pt}

\def \out{\leftskip = 0 pt\rightskip = 0pt}

\def\lf{16\pi^2}

{\nopagenumbers
\line{\hfil LTH 347}
\vskip .5in
\centerline{\titlefont Universality and the sparticle spectrum}
\vskip 1in
\centerline{\bf I.~Jack, D.R.T.~Jones and K.L.~Roberts}
\bigskip
\centerline{\it DAMTP, University of Liverpool, Liverpool L69 3BX, U.K.}
\vskip .3in
We calculate the particle spectrum of the SSM
which follows from the assumption
that the commonly assumed universal form of the soft \sy --breaking terms
is  invariant under renormalisation.
It is argued that this ``strong'' universality might be approached as an
infra--red fixed point for the unified theory above the
unification scale.
\Date{May 1995}}

\newsec{Introduction}

The unification of the gauge couplings at $M_{G}\approx 10^{16}$~GeV has
been responsible  for a much increased interest in the \ssm\ (SSM). It
is commonly assumed  that the \sy-breaking terms unify likewise, and so
are determined  ultimately by only four parameters: $m_0$, $A$, $B$ and
$M$,  which we will define presently. There have been many  attempts to
justify this {\it universal\/}  form for the soft breaking in terms of
$N=1$ supergravity, with or without an underlying string theory. In some
 scenarios the parameters turn out to be related, so that the soft
terms may be characterised by as few as one or two parameters.

At what scale does unification of the soft breakings take place? In view
of their gravitational  origin a first guess would place this  scale at
the Planck mass ($10^{19}$~GeV); most analyses, however, assume  that
soft unification holds, at least to a good approximation, at  the gauge
unification scale. In fact, in explicit models the soft  unification may
occur at some intermediate scale, but it seems  not unreasonable to
explore the consequences of locating it near or  at  $M_P$.   One may
then expect model--dependent deviations from universality  at $M_G$, and
the question is whether these deviations will  significantly impact low
energy predictions.   This program has been pursued recently in
Ref.~\ref\Pol {N.~Polonsky and A.~Pomarol, \prl 73 (1994) 2292.}, with
the conclusion that there can indeed be a quite substantial  effect on
the sparticle spectrum due to the evolution between  $M_P$ and $M_G$.

In a recent paper\ref\jjb{I.~Jack and D.R.T.~Jones, \plb 349 (1995)
294.},  two of us approached this issue from a different point of view.
We asked whether there existed any theories  such that universality (in
the sense described above) is a  renormalisation group invariant
property of the theory. Were the  unified effective field theory valid
for scales between $M_G$ and $M_P$ to have this property,  then clearly
universality at $M_P$ would imply universality at $M_G$.  We found,
remarkably, that this {\it strong\/}  universality is a property of  a
class of softly broken theories which satisfy one simple relation among
the  {\it dimensionless\/} coupling constants.  Moreover, the soft
breakings are all determined by the gaugino mass, $M$,  and their
relationships to each other bear intriguing similarity  to analogous
relations in certain string--based theories.  Even if the relations
implied by our constraints were not exact properties  of the fundamental
theory, they might still arise to a good approximation  in the
infra--red limit at $M_G$, from a more general  class of theories at
higher scales, and thus still be relevant at low  energies. In this
paper we pursue the phenomenological consequences of this idea,  for the
SSM.

\newsec{Universality}

The essential results from Ref.~\jjb\ are as follows.
We start with a \sic\ theory whose
 Lagrangian $L_{\rm SUSY} (W)$ is defined by the superpotential
\eqn\Aae{
W={1\over6}Y^{ijk}\Phi_i\Phi_j\Phi_k+{1\over2}\mu^{ij}\Phi_i\Phi_j.}

$L_{\rm SUSY}$ is the Lagrangian for  the $N=1$ supersymmetric
gauge theory, containing the gauge multiplet $\{A_{\mu},\lambda\}$
($\lambda$ being the
gaugino) and a chiral superfield $\Phi_i$ with component fields
$\{\phi_i,\psi_i\}$ transforming as a (in general reducible)
representation $R$ of the gauge group $\cal G$.
We assume for simplicity that there are no gauge-singlet
fields and that $\cal G$ is simple. (The generalisation to a semi-simple
group is trivial.)

The soft breaking is incorporated in $L_{\rm SB}$, given by
\eqn\Aaf{
L_{\rm SB}=(m^2)^j_i\Phi^{i}\Phi_j+
\left({1\over6}h^{ijk}\Phi_i\Phi_j\Phi_k+{1\over2}b^{ij}\Phi_i\Phi_j
+ {1\over2}M\lambda\lambda+{\rm h.c.}\right)}
(Here and elsewhere, quantities with superscripts are complex conjugates of
those with subscripts; thus $\Phi^i\equiv(\Phi_i)^*$.)

Our fundamental hypothesis is that the
 dimensionless couplings of the unified theory satisfy the
constraint
\eqn\Ak{
P^i{}_j={1\over3}g^2Q\delta^i{}_j.}
 where
\eqna\Aac$$\eqalignno{
Q&=T(R)-3C(G),\quad\hbox{and}\quad &\Aac a\cr
P^i{}_j&={1\over2}Y^{ikl}Y_{jkl}-2g^2C(R)^i{}_j. &\Aac b\cr}$$
Here
\eqn\Aaca{
T(R)\delta_{AB} = \Tr(R_A R_B),\quad C(G)\delta_{AB} =
f_{ACD}f_{BCD} \quad\hbox{and}\quad
C(R)^i{}_j = (R_A R_A)^i{}_j,}
where the $f_{ABC}$ are the structure constants of ${\cal G}$.

If we impose Eq.~\Ak, then the following relations among the soft breakings
are renormalisation group invariant through at least two--loops:
\eqna\Aj$$\eqalignno{h^{ijk}&=-MY^{ijk},&\Aj a\cr
(m^2)^i{}_j&={1\over3}(1-{1\over{\lf}}{2\over3}g^2Q)MM^*\delta^i{}_j,&\Aj b\cr
b^{ij}&=-{2\over3}M\mu^{ij}.&\Aj c\cr}$$
The fact that these relations are preserved under renormalisation
subject only to the simple constraint of Eq.~\Ak\ requires  a
miraculous sequence of cancellations among contributions from the various
$\beta$--functions; for a discussion, see Ref.~\jjb.

In the usual SSM  notation, Eq.~\Aj{}\ corresponds to
a universal scalar mass $m_0$  and universal $A$ and $B$ parameters
related (to lowest order in $g^2$) to the gaugino mass $M$ as follows:
\eqna\Baa$$\eqalignno{
 m_0 &= {1\over{\sqrt{3}}}M,&\Baa a\cr
A &= -M,&\Baa b\cr
B &= -{2\over3}M.&\Baa c\cr}$$

Remarkably,  relations of this form can arise in effective supergravity
theories motivated by superstring theory, where \sy\ breaking is assumed to
occur purely via vacuum expectation values for dilaton  and moduli fields
\ref\IL{L.E. Ib\'a\~nez and D. L\"ust,
\npb382 (1992) 305\semi
V. Kaplunovsky and J. Louis, \plb306 (1993) 269}
\ref\CM{A. Brignole, L. E. Ib\'a\~nez and C. Mu\~noz, \npb422 (1994) 125;
erratum--{\it ibid} 436 (1995) 747.}.
Ignoring string loop corrections and possible phases for the
auxiliary fields $F^S$ and $F^T$, where $S$ is the dilaton and $T$ is the
overall modulus, then according to Ref.~\CM,

\eqna\Keva$$\eqalignno{
m^2_0 &= {1 \over 3} M^2 + {2 \over 3} {M^2 \over C^2 \sin^2\theta}
(C^2 -1) &\Keva a\cr
A &= -M &\Keva b\cr
}$$
and
\eqn\Kevb{
B=-{M(1+\sqrt3\sin\th+C\cos\th)\over{\sqrt3 C\sin\th}}}
or
\eqn\Kevc{
B={2M\over{\sqrt3 C\sin\th}},
}
where Eqs.~\Kevb\ and \Kevc\ apply according to whether
the $\mu$--term is generated by an explicit $\mu$--term in
the supergravity superpotential, or by a special term in the K\"ahler
potential.

Here $C$ is related to the vacuum expectation
value of the scalar potential and a vanishing cosmological constant
corresponds to $C =1$. $\theta$ is called the {\it goldstino mixing
angle}, and the values $\th =0$ and $\th =\pi /2$ correspond to
modulus--dominated and dilaton--dominated  cases respectively.
It is easy to see that
with $C=1$, Eqs.~\Keva{a}\ and \Keva{b}\
reproduce Eqs.~\Baa{a}\ and \Baa{b},  and  for $\theta =
4\pi / 3$, Eq.~\Kevb\ gives Eq.~\Baa{c}.

Another particular case that has been subject to some phenomenological
investigation
\ref\blm{R.~Barbieri, J.~Louis and M.~Moretti, \plb 312 (1993) 451;
erratum--{\it ibid\/} 316 (1993) 632.}
\ref\lopez{J.L.~Lopez, D.V.~Nanopoulos and A.~Zichichi, \plb 319 (1993) 451.}
is that of $C=1$ and $\theta = \pi /2$ in Eq.~\Kevc .
We will refer to this case as the $DD$ (dilaton--dominated) scenario.
It again
corresponds to Eq.~\Baa{}\ except that now $B = 2M/\sqrt{3}$.
We shall
see that this difference has considerable impact.
The similarity between the conditions on the soft-breaking terms
which arise from our strong universality hypothesis and those that emerge
from string theory is certainly intriguing.

There is, however, an alternative interpretation of the above results.
Consider a unified theory where it would be {\it possible\/} to find
Yukawa couplings satisfying  Eq.~\Ak. The fact that Eqs.~\Ak\
and~\Baa{}\  are renormalisation group invariant is of course equivalent
to saying that they are  fixed points of the evolution equations; fixed
points, moreover, that are approached in the infra--red, at least in
simple examples. At first sight it might appear that the difference
between $M_P$  and $M_G$ is insufficient to allow significant evolution,
but it has recently  been argued\ref\ross {M.~Lanzagorta and G.G.~Ross,
\plb 349 (1995) 319.} that in the case of the Yukawa couplings the
evolution towards the fixed point occurs more rapidly  in the unified
theory than in the low energy theory. If we believe that this
conclusion holds also for the soft terms, then it is possible to argue
that for a wide range of  input parameters the boundary conditions  of
Eq.~\Baa{}\ might hold at $M_G$.

Our philosophy now is as follows. We assume that the SSM  is valid
below gauge unification, and that the unified theory satisfies Eq.~\Ak.
We then proceed to impose Eq.~\Baa{a-c} as boundary conditions at
the gauge unification scale. These boundary conditions are so restrictive
that it is not a priori obvious that a phenomenologically viable solution
will exist.

\newsec{The \ssm}

We start with the superpotential:
\eqn\ssma{
W = \mu_s H_1 H_2 + \lambda_t H_2 Q\tbar  + \lambda_b H_1 Q\bbar
+ \lambda_{\tau} H_1 L\taub}
where we neglect Yukawa couplings except for those of the third
generation.

The Lagrangian for the SSM  is  defined by the superpotential of
Eq.~\ssma\ augmented with  soft breaking terms as follows:
\eqn\ssme{L_{SSM} = L_{\rm SUSY}(W) + L_{\rm SOFT}}
where
\eqn\ssmf{\eqalign{
L_{\rm SOFT} = &-m_1^2 H_1^{\dag}H_1 - m_2^2 H_2^{\dag}H_2
+ [m_3^2 H_1 H_2  + \hbox{h.c.} ]\cr
& -\sum_i \left( m_Q^2 |Q|^2 + m_L^2 |L|^2
+m^2_{\tbar} |\tbar |^2 +m^2_{\bbar} |\bbar |^2
+ m^2_{\taub} |\taub |^2\right)\cr
&+ [ A_t\la_t H_2 Q\tbar + A_b \la_b H_1 Q\bbar +
A_{\tau}\la_{\tau} H_1 L\taub  +  \hbox{h.c.} ]\cr
&-\half [ M_1\la_1\la_1 + M_2\la_2\la_2 + M_3\la_3\la_3 +  \hbox{h.c.} ]\cr}}
and the sum over $i$ for the $m^2$ terms is a sum over the three generations.

The running coupling  and mass analysis of the above theory has been performed
many times. The novel feature here is the very restricted set of boundary
conditions at gauge unification:
\eqna\ssmg$$
\eqalignno{&m_1 = m_2 = m_Q = m_L = m_{\tbar} = m_{\bbar}
= m_{\taub} = {1\over{\sqrt{3}}}M, &\ssmg a\cr
&A_{\tau} = A_b = A_t = -M, &\ssmg b\cr
&M_1 = M_2 = M_3 = M, &\ssmg c\cr
&m_3^2 = -{2\over 3}\mu_s M, &\ssmg d\cr}
$$
where Eq.~\ssmg{a}\ includes the squarks and sleptons of all
three generations.
Notice that these boundary conditions satisfy the constraint
\eqn\ssmga{
 \Delta = m_1^2 + m_2^2 + 2\mu_s^2 - 2|m_3^2| > 0
}
for any $\mu_s$.  We require this (at $M_G$)
to keep the  potential bounded from below, in other words so that
$SU_2\otimes U_1$ breaking does not occur with characteristic scale $M_G$.
(See Eq.~[A.2]. It is interesting that in the $DD$ scenario, one  obtains
\eqn\ssmgb{\Delta = 2\left({M\over{\sqrt3}}-\mu_s \right)^2.}
With
Eq.~\Kevc\ and a value
of the goldstino angle $\th$ other than ${\pi\over 2}$, one would need
to check that $\mu_s (M_G)$ indeed gave $\Delta > 0$.)

Our procedure is as follows. We input $\alpha_1$, $\alpha_2$,
$\alpha_3$,  $m_t$ and $\tan\beta$ at $M_Z$, and  calculate the
unification scale  $M_G$ (defined as the meeting point of  $\alpha_1$
and $\alpha_2$) by running the dimensionless couplings. Then we input
the gaugino mass  $M$ at $M_G$, and run the dimensionful parameters
(apart from $m_3^2$ and $\mu_s$)  down to $M_Z$. We can then determine
$m_3^2$ and $\mu_s^2$ as usual  at $M_Z$ by minimising  the (one--loop
corrected) Higgs potential. Then we run $m_3^2$ and $\mu_s$ back  up to
$M_G$ (for the two possibilities of $\hbox{ sign }\,\mu_s$)  and calculate
$B' = B/M = (m_3^2)/(M\mu_s)$. By plotting $B'$ against the input  value
of $\tan\beta$ we can then determine whether (for a given input $M$)
there  exists a value of $\tan\beta$ such that Eq.~\ssmg{d}\ is
satisfied.   Given a set $M, \tan\beta$ satisfying our boundary
conditions we  can calculate the sparticle spectrum in the usual way and
plot the resulting masses against $M$. See the appendix for some
comments about the $\beta$--functions and mass matrices.

At this stage we are chiefly interested in demonstrating that
phenomenologically
viable solutions are possible with
our highly restricted boundary conditions. Consequently we ignore
threshold corrections to the mass predictions (for a recent discussion of
threshold corrections  see,
for example, Ref.~\ref\bagg{J.~Bagger, K.~Matchev and D.~Pierce,
hep-ph/9501277.}). Nor do we address here
the recent concerns
\bagg\ref\shifman{L.~Roszkowski and M.~Shifman, hep-ph/9503358.} relating
to the apparent incompatibility (in a precision calculation) of
the experimental value of $\alpha_3 (M_Z)$ and the value required
for gauge unification (note that the solution proposed in
Ref.~\shifman, to wit non--unified gaugino masses, is not available to us).

 We do, however,
incorporate  the one--loop corrections into the minimisation of the
Higgs potential.\footnote{\dag}{The necessity for doing this was first
pointed out in Ref.~\ref\Gamb {G.~Gamberini, G.~Ridolfi and F.~Zwirner,
\npb 331 (1990) 331.}}In general we have done this by
solving the Higgs tadpole equations,
but we also checked our results by numerically minimising the
Higgs potential in some cases. (Our results for the Higgs tadpoles
agree with Ref.~\ref\bbo{V.~Barger, M.S.~Berger and  P.~Ohmann,
\prd 49 (1994) 4908.}, apart from one or two minor typos.)
We also do  include one loop corrections to the
mass ($m_h$) of the lighter  CP--even Higgs  boson, since, as is well
known, the radiative corrections  are important in
this case\ref\haber{H.E.~Haber and R.~Hempfling, \prl 66
(1991) 1815; \prd 48 (1993) 4280; Y.~Okada, M.~Yamaguchi and T.~Yanagida,
Prog. Theor. Phys. {\bf 85}
(1991) 1; \plb 262 (1991) 54;
J.~Ellis, G.~Ridolfi and F.~Zwirner, \plb
257 (1991) 83; {\it ibid\/} 262 (1991) 477;
P.H.~Chankowski, S.~Pokorski and J.~Rosiek, \plb 274 (1992) 191;
{\it ibid\/} 281 (1992) 100;
R.~Barbieri and M.~Frigeni, \plb 258 (1991) 395;
A.~Yamada, \plb 263 (1991) 233;
A.~Brignole, \plb 281 (1992) 284;
M.~Drees and M.M.~Nojiri, \prd 45 (1992) 2482
.}. More precisely,  our results for $m_h$ and $m_H$
are based on the appropriate second
derivative  of the one--loop corrected effective potential evaluated
with the  scale $\mu$ set equal to the gaugino mass $M$. While this is
crude compared  to existing calculations, it incorporates the most
important  logarithmic effects. Our results for other masses are based
on the tree  mass matrices but again  with all running parameters
evaluated at the scale $M$.

Since  the two--loop corrections to the $\beta$--functions are now available
\ref\jja{I.~Jack and D.R.T.~Jones, \plb 333 (1994) 372.}%
\nref\mv{S.P.~Martin and M.T.~Vaughn, \prd50 (1994) 2282.}%
\nref\yamada{Y.~Yamada, \prd50 (1994) 3537.}%
--\ref\jjmvy{I.~Jack,
D.R.T.~Jones, S.P.~Martin, M.T.~Vaughn  and Y.~Yamada, \prd 50 (1994) R5481.},
we incorporate these. In general
their effect is very small, being most noticeable in the Higgs sector;
although the mass of the lightest Higgs is essentially
unchanged, the other Higgs masses are increased by up to $10\%$
by the two loop corrections.   Of course for precise predictions,
we should also include threshold corrections, as indicated above.

We use input parameters at $M_Z$ as follows:
\eqn\ssmh{\eqalign{
\alpha_1 &= 0.0169, \quad \alpha_2 = 0.0337, \quad \alpha_3 = 0.11\cr
m_{\tau}(M_Z) &= 1.75 \hbox{ GeV},\quad  m_b (M_Z) = 3.2\hbox{ GeV},
\quad m_t (M_Z) = 170\rightarrow 200\hbox{ GeV}.\cr}}

The appropriate input $m_b (M_Z)$ depends itself on the
sparticle spectrum in general, as emphasised recently
\ref\hall{R.~Hempfling, \prd 49 (1994) 6168;
B.D.~Wright, MAD/PH/812, hep--ph/9404217;
L.J.~Hall, R. Rattazzi and U. Sarid, \prd 50 (1994) 7048;
hep--ph/9405313;
M.~Carena, M.~Olechowski,
S.~Pokorski and C.E.M.~Wagner, \npb 426 (1994) 269.}.
Varying $m_b (M_Z)$ does not, however,
affect the qualitative nature of our results.

\newsec{Discussion of the results: ${\bf m_t= 175\hbox{ GeV}}$}

In this section we consider in detail the case $m_t (M_Z) = 173$~GeV
which corresponds to a pole mass $m_t \approx 175$~GeV.

Fig.~(1) plots $B'$ against $\tan\beta$ for $M=200$~GeV and
 a pole top mass of $175$~GeV.
\medskip
\epsfysize= 4.0in
\centerline{\epsffile{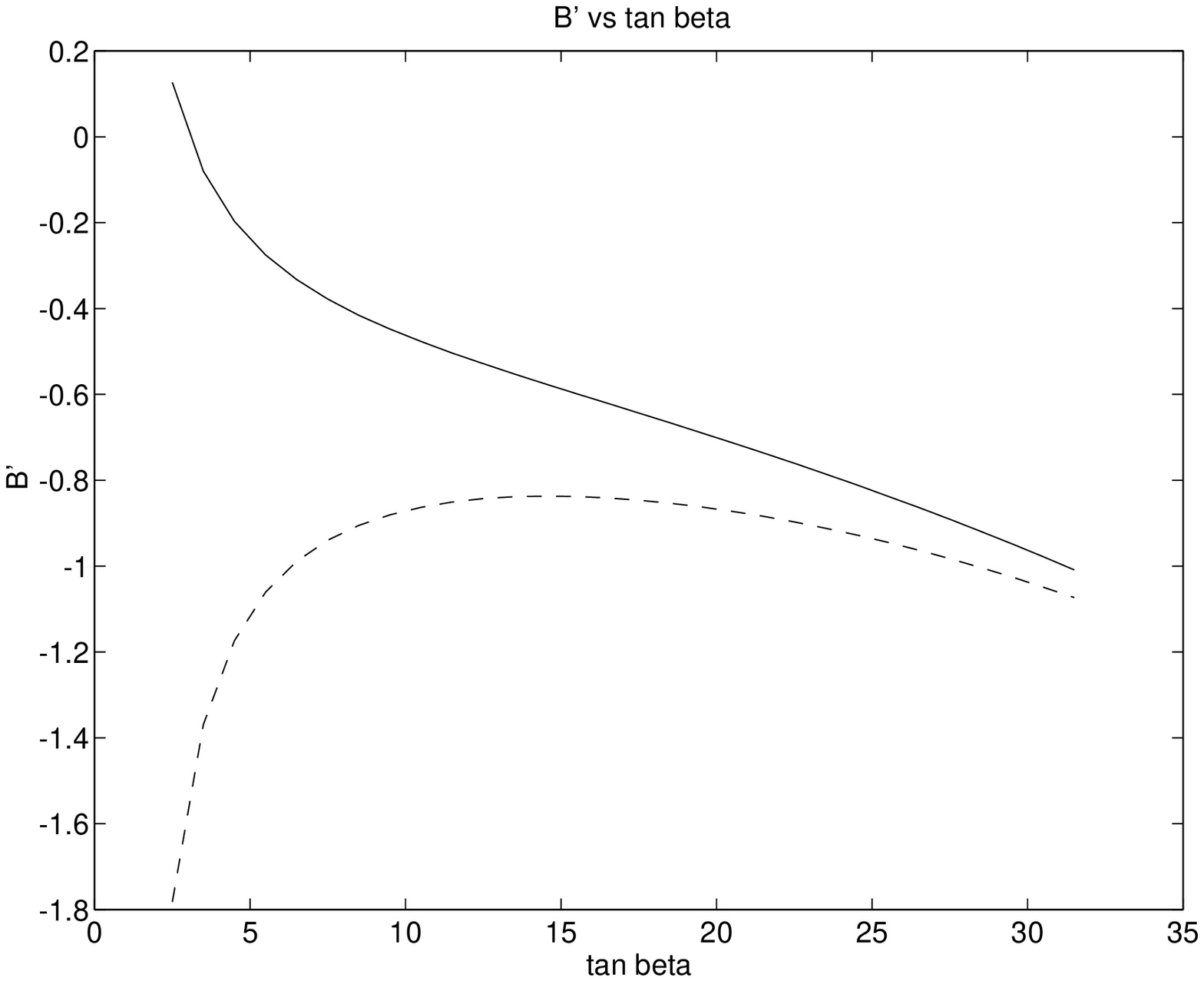}}
\in
{\it \noindent Fig.1: The $B'$-parameter vs.
$\tan\beta$ for input gaugino mass $M=200$~GeV
and $m_t = 175$~GeV. The solid and dashed lines correspond to $\mu_s > 0$ and
$\mu_s < 0$ respectively.
The required value of $B'$ is obtained for $\tan\beta\approx 18$.
}
\medskip
\out
Now since we want $B' = -{2\over 3}$, we might expect (with our conventions)
to find a solution with $\mu_s<0$ rather than $\mu_s>0$.
(This is because for a tree minimum at the weak scale we necessarily
would have $m_3^2 > 0$.)
 We see, in fact, that with $\mu_s<0$ we do  indeed get $B' < 0$
throughout;  but unfortunately,  $B' = -{2\over 3}$ is not
possible for any $\tan\beta$.  Surprisingly, the  situation is better
with $\mu_s > 0$, and we have the desired result for
$\tan\beta \approx 18$.  For the $DD$ scenario, when the desired result is
$B' = {2\over \sqrt{3}}$, notice that the solution (had it existed)
would have been for $\mu_s>0$ and in the small $\tan\beta$ region. This
solution is  vulnerable to the
existence of the well--known Landau pole in the top mass Yukawa, at
$m_t \approx 195\sin\beta$. Thus Fig.~(1) is consistent with
the conclusions of Ref.~\lopez, which quotes an upper limit
on $m_t$ of $155$~GeV for the $DD$ case.  (Note that Ref~\lopez\ has the
opposite sign for $\mu_s$.) This means that the strict $DD$ scenario
is ruled out by the recent measurements of $m_t$
\ref\topmass{F.~Abe et al., \prd 50 (1994) 2966; hep-ex/9503002;
S.~Abachi et al., hep-ex/9503003.}, though of course the
general string--based framework for the origin of the soft terms,
in which $B'$ is a free parameter, is not compromised.

In Fig.~(2) we plot $\tan\beta$ against the input gaugino mass, $M$,
having satisfied Eq.~\ssmg{d}. ($M$ is related to the gluino
mass $M_g (M_Z)$, by the relation $M_g(M_Z)\approx 2.4 M$, but
note that, especially for large $M$, the gluino pole mass
can differ considerably from $M_g (M_Z)$).
\medskip
\epsfysize= 4.0in
\centerline{\epsffile{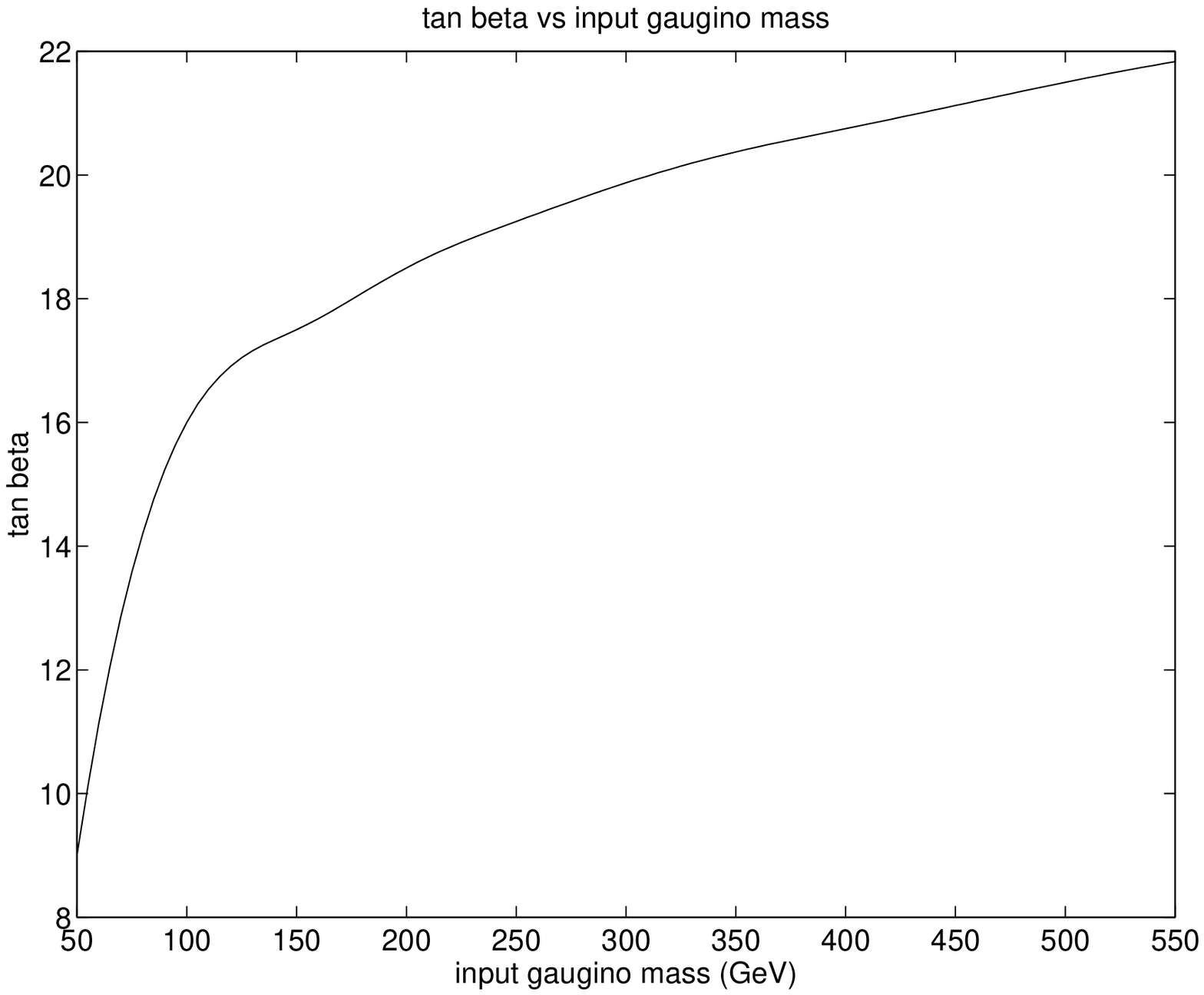}}
{\it{\centerline{Fig.~2:   $\tan\beta$ vs. $M$ for  $m_t = 175$~GeV.}}}
\medskip
As already mentioned, we find comparatively large values of $\tan\beta$ ,
except for small $M$. As is well known, successful bottom--tau
Yukawa unification favours a large top Yukawa coupling
\ref\bjd{J.E.~Bj\"orkman and D.R.T.~Jones, \npb 259 (1985) 533.}
\ref\barger{V.~Barger, M.S.~Berger and P.~Ohmann, \prd 47 (1993)1093;
M.~Carena, S.~Pokorski and C.E.M.~Wagner, \npb 406 (1993) 59;
 C.~Kolda, L.~Roszkowski, J.D.~Wells and G.L.~Kane,
\prd 50 (1994) 3498.}, and so we do not obtain it within our approach,
at least within our approximation. This conclusion is sensitive to
the input $m_b (M_Z)$, which is in turn affected by radiative corrections
(especially that involving the gluino) which will not be negligible
for $\tan\beta\approx 18$. At first sight, however, these corrections
take us further from $b$--$\tau$ unification. This point deserves further
investigation, but we are in any case not too concerned,
however, since $b$--$\tau$ unification is a model dependent
phenomenon.

Fig.~(3) plots the Higgs masses against
the gaugino mass.\footnote{*}{In Fig.~3 (and Figs.~4, 5)
$\tan\beta$ changes with the gaugino mass
in accordance with Fig.~2.}
\medskip
\epsfysize= 4.0in
\centerline{\epsffile{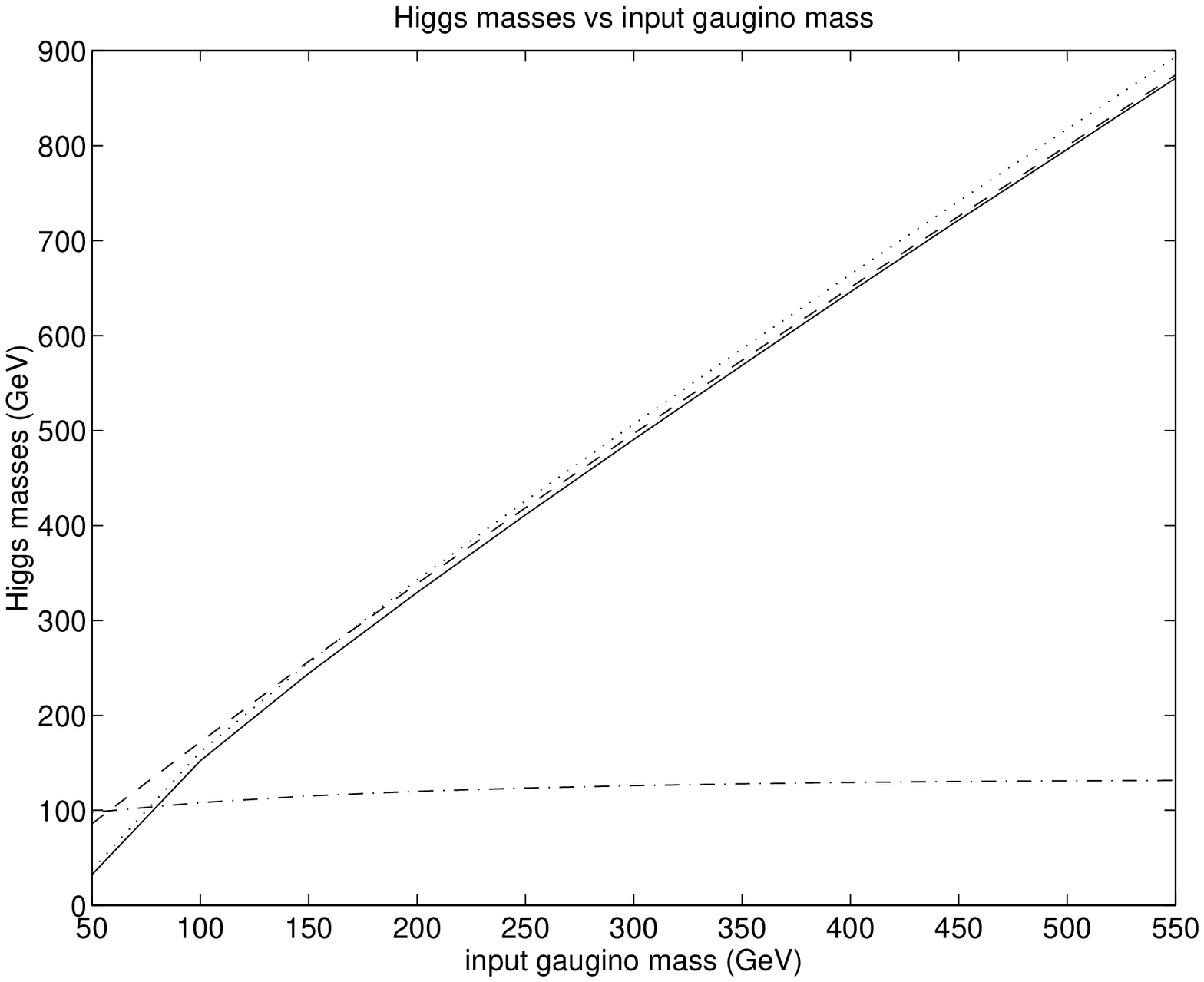}}
\in
{\it \noindent Fig.3: The Higgs masses vs. $M$ for
  $m_t = 175$~GeV. The
solid and dashed lines  correspond
to the approximately degenerate $CP$-odd neutral and
charged Higgses respectively. The dotted line is $m_H$ and the dot--dashed
line is $m_h$.}
\medskip
\out
The one--loop radiative corrections,
which we have included, raise $m_h$ above the tree bound
$m_h < | M_Z\cos 2\beta |$.
Our result for $m_h$ is dependent only weakly on  the input gaugino mass,
with $m_h\approx 115$~GeV. This is consistent with the generally
accepted bound $m_h\leq 135$~GeV (or $m_h\leq 146$~GeV in more general
models\ref\kane{H.~Haber and M.~Sher, \prd 35 (1987) 2206;
J. Gunion, L. Roszkowski and H. Haber, \plb 189 (1987) 409;
\prd 38 (1988) 105;
M. Drees, \ijmpa 4 (1989) 3635;
P.~Bin\'etruy and C.~Savoy, \plb 277 (1992) 453;
J.R.~Espinosa and M.~Quiros, \plb 279 (1992) 92;
 G.L.~Kane, C.~Kolda and J.D.~Wells, \prl 70 (1993) 2686.}).

For $M=150$~GeV we have $M_{h, H} = 116, 257$~GeV and
$m_{A, H^{\pm}} = 246, 259$~GeV. It is interesting to compare these
results with those
obtained if one--loop $\beta$--functions are used throughout, which
are $M_{h, H} = 116, 242$~GeV and
$m_{A, H^{\pm}} = 225, 239$~GeV; so the corrections to $m_{A, H^{\pm}}$
are $O(10\%)$. The masses of the sparticles
are in general less affected by using two--loop rather than one--loop
$\beta$--functions; typically a sparticle mass changes by $5\%$ or so.

Fig.~(4) plots the neutralino masses against the gaugino mass.
\bigskip
\epsfysize= 4.0in
\centerline{\epsffile{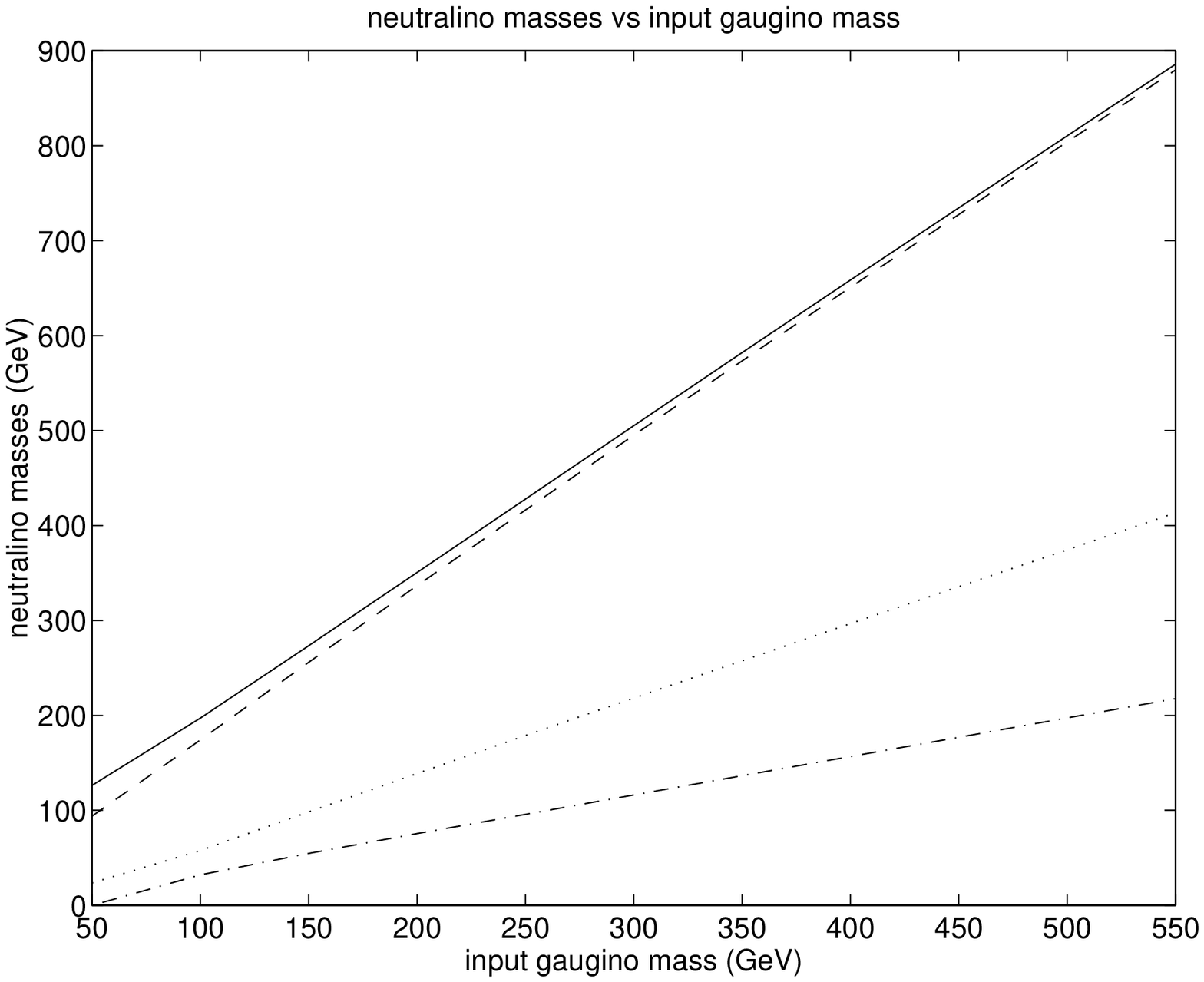}}
\in
{\it \noindent Fig.4: The neutralino masses vs. $M$ for
  $m_t = 175$~GeV. The
solid and dashed lines correspond to the Higgsino--dominated
neutralinos, and the  dotted and  dot--dashed
lines to the gaugino--dominated neutralinos.}
\medskip
\out
Except for small gaugino masses, the
lightest neutralino is the lightest superpartner. For $M = 150$ GeV,
for example, we have $m^0_{\chi_{4}} \approx 55$~GeV which is
potentially interesting as cold dark matter. Of course the precise
$\chi$ relic density  is controlled by the $\chi\chi$
annihilation  cross--section, so we need to investigate this to
test this hypothesis. (For a review of particle physics dark matter
candidates, see, for example, Ref.\ref\ellis{J.~Ellis, Les Houches lectures,
CERN-TH.7083/93.}).
Fig.~(5) plots the $\tau$--slepton masses against the gaugino mass.

\bigskip
\epsfysize= 4.0in
\centerline{\epsffile{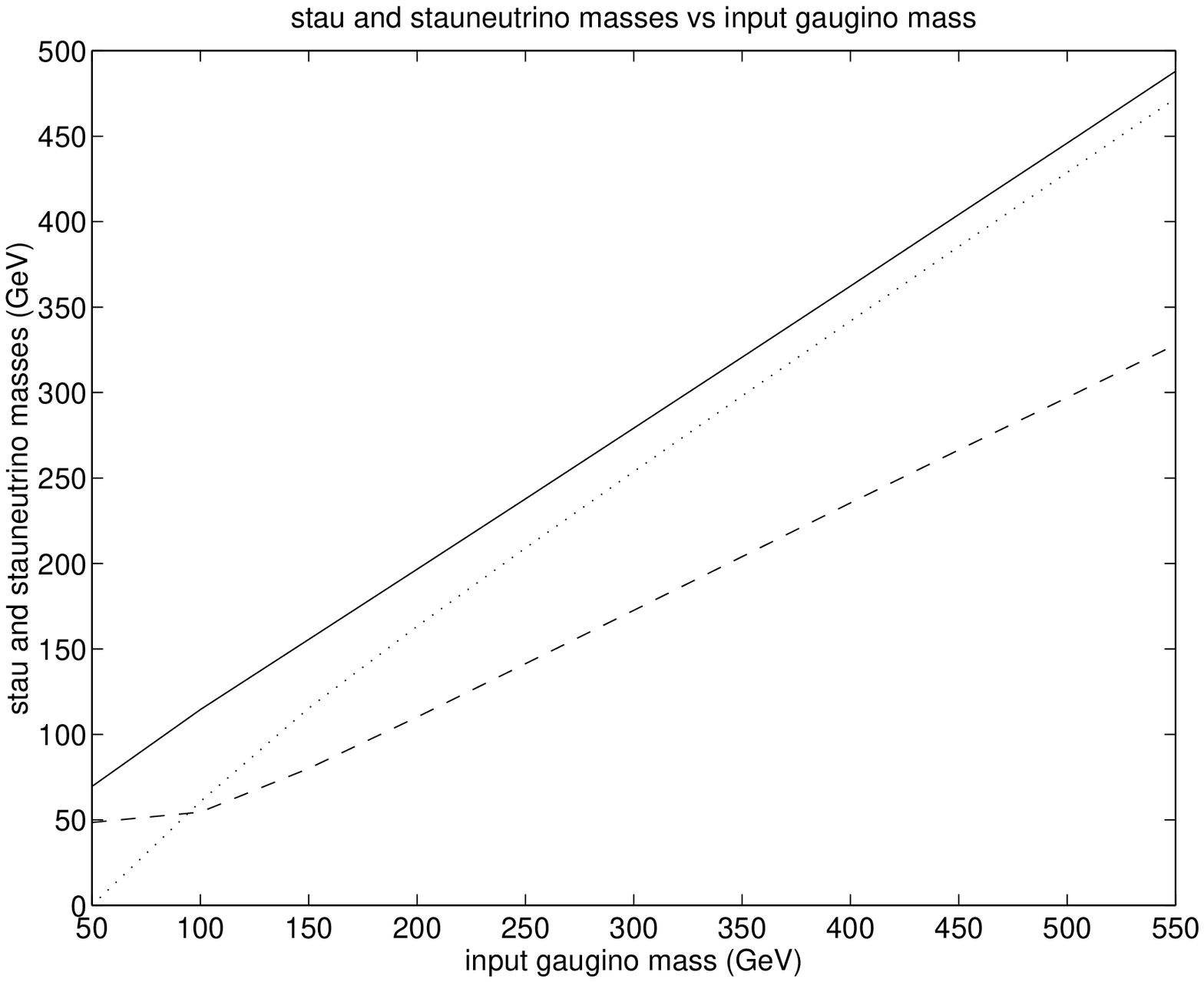}}
\in
{\it \noindent Fig.5: The $\tau$-slepton masses vs. $M$ for
  $m_t = 175$~GeV. The
solid and dashed lines correspond to $\tautilde_{1,2}$ and
the  dotted line is the $\nutilde_{\tau}$}.
At $M=150$~GeV we have $M_{\tautilde_{1,2}} \approx 156, 80$~GeV.
\medskip
\out

It will be apparent that the  plots presented thus far  exhibit
linear behaviour for a wide range of input gaugino masses.
Rather than give more figures, we therefore summarise our
results in Table~1, which
gives a good approximation  (within a few GeV) for
$100\hbox{ GeV }< M < 500\hbox{ GeV}$.

 With $m_t=185$~GeV, the
dependence of $B'$ on $\tan\beta$ and the resulting
sparticle spectrum are very similar. For $m_t\geq 190$~GeV, there is a
change, which we discuss in the next section; but we give results
for $m_t = 190$~GeV here as well, for simplicity.

$$\vbox{\offinterlineskip
\def\vr{\vrule height 11pt depth 5pt}
\def\vrq{\vr\quad}
\settabs
\+
\vrq 1.0\quad & \quad 1.0 \quad & \vrq 1.0 \quad & \quad 1.0 \quad &
\vrq 1.0 \quad
& \quad 1.0 \quad &
\vrq 1.0 \quad & \quad 1.0 \quad &
\vr\cr\hrule
\+
\vrq $m_t$ && \vrq 175 && \vrq 185 && \vrq 190 &&\vr\cr\hrule
\+
\vrq $m = aM + b$ &&
\vrq $a$ \quad & \vrq $b$ \quad &
\vrq $a$ \quad & \vrq $b$ \quad &
\vrq $a$ \quad & \vrq $b$ \quad &
\vr\cr\hrule
\+
\vrq $m_h$ &&
\vrq 0.048 \quad & \vrq 108 \quad &
\vrq 0.059 \quad & \vrq 108 \quad &
\vrq 0.070 \quad & \vrq 106 \quad &\vr\cr\hrule
\+
\vrq $m_H$ && \vrq 1.613 \quad & \vrq 15 \quad & \vrq 1.800 \quad &
\vrq 7 \quad &\vrq 1.870 \quad & \vrq 5 \quad &\vr\cr\hrule
\+
\vrq $m_A$ && \vrq 1.585 \quad & \vrq  8 \quad & \vrq 1.782 \quad &
\vrq 4 \quad &\vrq 1.855 \quad & \vrq 2 \quad &\vr\cr\hrule
\+
\vrq $m_{H^{\pm}}$ && \vrq 1.555 \quad & \vrq 25 \quad & \vrq 1.755 \quad &
\vrq 20 \quad &\vrq 1.829 \quad & \vrq 17 \quad &\vr\cr\hrule
\+
\vrq $m_{\etilde_1}$ && \vrq 0.872 \quad & \vrq 12 \quad & \vrq 0.873 \quad &
\vrq 12 \quad &\vrq 0.874 \quad & \vrq 11 \quad &\vr\cr\hrule
\+
\vrq $m_{\etilde_2}$ && \vrq 0.666 \quad & \vrq 12 \quad & \vrq 0.667 \quad &
\vrq 12 \quad &\vrq 0.668 \quad & \vrq 12 \quad &\vr\cr\hrule
\+
\vrq $m_{\nutilde_e}$ && \vrq 0.930 \quad & \vrq -22 \quad & \vrq 0.930 \quad &
\vrq -21 \quad &\vrq 0.930 \quad & \vrq -21 \quad &\vr\cr\hrule
\+\vrq $m_{\tautilde_1}$ && \vrq 0.830 \quad & \vrq 31
\quad & \vrq 0.852 \quad &
\vrq 22 \quad &\vrq 0.861 \quad & \vrq 18 \quad &\vr\cr\hrule
\+\vrq $m_{\tautilde_2}$ && \vrq 0.615 \quad & \vrq -11 \quad &
\vrq 0.644 \quad &
\vrq 1 \quad &\vrq 0.657 \quad & \vrq 5 \quad &\vr\cr\hrule
\+\vrq $m_{\nutilde_{\tau}}$ && \vrq 0.903 \quad &
\vrq -21 \quad & \vrq 0.917 \quad &
\vrq -21 \quad &\vrq 0.923 \quad & \vrq -20 \quad &\vr\cr\hrule
\+\vrq $m_{\chi_1^+}$ && \vrq 1.527 \quad & \vrq 48 \quad & \vrq 1.580 \quad &
\vrq 46 \quad &\vrq 1.601 \quad & \vrq 45 \quad &\vr\cr\hrule
\+\vrq $m_{\chi_2^+}$ && \vrq 0.793 \quad & \vrq -21 \quad & \vrq 0.799 \quad &
\vrq -23 \quad &\vrq 0.805 \quad & \vrq -25 \quad &\vr\cr\hrule
\+\vrq $m_{\chi_1^0}$ && \vrq 1.532 \quad & \vrq 44 \quad & \vrq 1.583 \quad &
\vrq 44 \quad &\vrq 1.603 \quad & \vrq 45 \quad &\vr\cr\hrule
\+\vrq $m_{\chi_2^0}$ && \vrq 1.566 \quad & \vrq 22 \quad & \vrq 1.622 \quad &
\vrq 20 \quad &\vrq 1.645 \quad & \vrq 18 \quad &\vr\cr\hrule
\+\vrq $m_{\chi_3^0}$ && \vrq 0.789 \quad & \vrq -19 \quad & \vrq 0.793 \quad &
\vrq -20 \quad &\vrq 0.797 \quad & \vrq -21 \quad &\vr\cr\hrule
\+\vrq $m_{\chi_4^0}$ && \vrq 0.410 \quad & \vrq -7 \quad & \vrq 0.413 \quad &
\vrq -8 \quad &\vrq 0.417 \quad & \vrq -9 \quad &\vr\cr\hrule
\+\vrq $m_{\utilde_1}$ && \vrq 2.264 \quad & \vrq 26 \quad & \vrq 2.266 \quad &
\vrq 26 \quad &\vrq 2.269 \quad & \vrq 26 \quad &\vr\cr\hrule
\+\vrq $m_{\utilde_2}$ && \vrq 2.189 \quad & \vrq 29 \quad & \vrq 2.191 \quad &
\vrq 30 \quad &\vrq 2.194 \quad & \vrq 30 \quad &\vr\cr\hrule
\+\vrq $m_{\dtilde_1}$ && \vrq 2.245 \quad & \vrq 37 \quad & \vrq 2.247 \quad &
\vrq 37 \quad &\vrq 2.251 \quad & \vrq 37 \quad &\vr\cr\hrule
\+\vrq $m_{\dtilde_2}$ && \vrq 2.175 \quad & \vrq 33 \quad & \vrq 2.177 \quad &
\vrq 33 \quad &\vrq 2.180 \quad & \vrq 33 \quad &\vr\cr\hrule
\+\vrq $m_{\ttilde_1}$ && \vrq 1.829 \quad & \vrq 143 \quad & \vrq 1.849 \quad
&
\vrq 143 \quad &\vrq 1.861 \quad & \vrq 142 \quad &\vr\cr\hrule
\+\vrq $m_{\ttilde_2}$ && \vrq 1.645 \quad & \vrq 0 \quad & \vrq 1.615 \quad &
\vrq 18 \quad &\vrq 1.609 \quad & \vrq 27 \quad &\vr\cr\hrule
\+\vrq $m_{\btilde_1}$ && \vrq 2.040 \quad & \vrq 56 \quad & \vrq 2.113 \quad &
\vrq 46 \quad &\vrq 2.142 \quad & \vrq 42 \quad &\vr\cr\hrule
\+\vrq $m_{\btilde_2}$ && \vrq 1.963 \quad & \vrq 20 \quad & \vrq 1.992 \quad &
\vrq 28 \quad &\vrq 2.009 \quad & \vrq 30 \quad &\vr\cr\hrule
}$$
\in
{\it \noindent Table 1: Linear approximations of the form  $m = aM +b$
to the mass spectrum for
  $m_t = 175$~GeV, $m_t = 185$~GeV and $m_t = 190$~GeV.}

\out
We will not perform a detailed analysis of our predictions vis-a-vis
current experimental limits; in
more general cases many treatments exist (for a recent example,
see Ref.~\ref\drees{M.~Drees and S.P.~Martin, MADPH-95-879, (hep-ph/9504324).})
It is clear enough that these
will impose a lower bound on $M$ of around $100$~GeV, and
that for say, $M\approx 150$~GeV we have acceptable
phenomenology, with a  stable neutralino at 55GeV,
a $\tau$-slepton at 80GeV, and the light Higgs at $115$~GeV.

\newsec{The large $\bf{m_t}$ region}

For large top masses (in the region $m_t\geq 190$~GeV) the nature
of the solutions we find to our universality constraints changes
in an interesting way. We find that for $\mu_s<0$ the dependence
of $B'$ on $\tan\beta$ ceases to be monotonic and that for a given
input gaugino mass there may be three possible values of $\tan\beta$
that give $B' = -{2\over 3}$. This behaviour is shown in Fig. 6,
for $M= 150$~GeV.

\epsfysize= 3.5in
\centerline{\epsffile{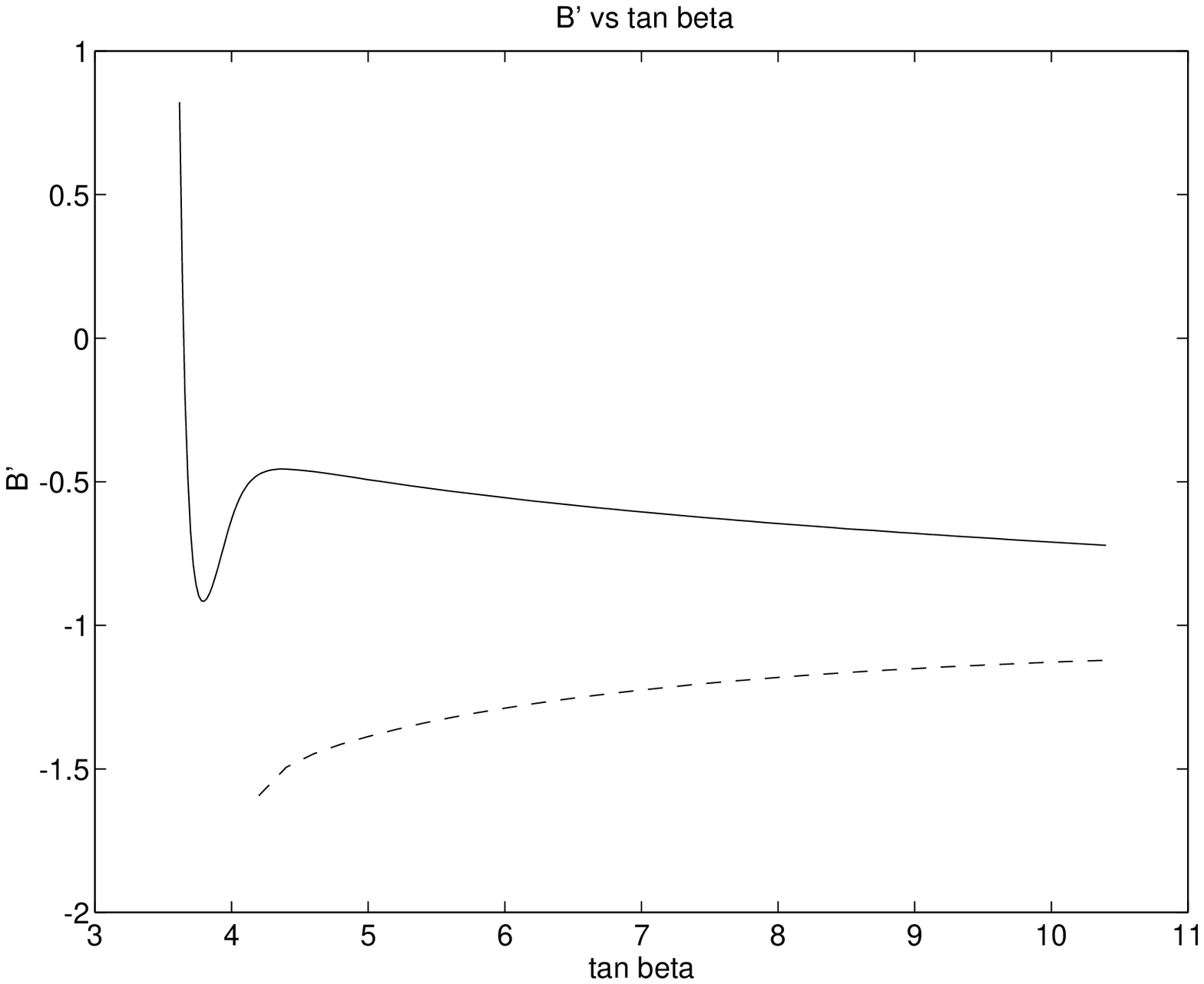}}
\in
{\it \noindent Fig.6: The $B'$-parameter vs.
$\tan\beta$ for input gaugino mass $M=200$~GeV
and $m_t = 190$~GeV. The solid and dashed lines correspond to $\mu_s > 0$ and
$\mu_s < 0$ respectively.
}
\medskip
\out

In fact, however, the existence of the
two solutions at $\tan\beta\approx 3.6\hbox{ and } 4.0$
depends on our use of the two--loop $\beta$--functions for the
dimensionless couplings; if we use the corresponding one loop ones they
do not exist because of the Landau pole in the top Yukawa coupling.
They are therefore unreliable, and we ignore them.
For the solution at $\tan\beta\approx 8$, the spectrum is similar to that
described in the last section, and is shown in Table~1, in the previous
section. For $m_t > 195$~GeV we are unable to satisfy Eq.~\Baa{c}\
and retain  perturbative unification.

\newsec{Conclusions}

We have shown that the restrictions imposed by the conjecture
of strong universality at $M_G$ leave a viable and
well determined
\sic\ phenomenology. The main new  feature of the
resulting spectrum is the determination (for given input
gaugino mass $M$) of $\tan\beta$. Although this occurs also in the $DD$
approach, the results for the two cases are readily
distinguished.   Since (given $m_t$)
the mass spectrum depends on a single parameter, $M$, it is clear that
the discovery of \sic\ particles would swiftly decide whether
our marriage of universality with the minimal SSM  corresponds
to reality.

It would obviously be nice if we could construct a unified theory that
satisfied (or approached in the infra--red limit at $M_G$) our
strong  universality hypothesis as encapsulated in Eq.~\Ak and ~\Aj .
In this connection, it is worth observing that Eq.~\Ak\ permits
gauge groups with $U_1$ factors (in contrast to the finite
case, $P = Q = 0$). Then the conditions Eq.~\Aj{}\ still suffice for a
universal theory as long as we have also that
\eqn\cona{
\hbox{ Tr }(R_A m^2) = m^2\hbox{ Tr }R_A = 0}
which is the condition that the theory be free of gravitational anomalies
\ref\gaume{L.~Alvarez--Gaum\'e and E.~Witten, \npb 234 (1984) 269.}.
In the light of this remark, a theory based on ``flipped'' $SU_5$
($SU_5\otimes U_1$ -- see for example Ref~\ref\flipped{
J.L.~Lopez, D.V.~Nanopoulos and Ka-jia
Yuan, \npb 399 (1993) 654.})
 might be worth a try, though the direct product
nature of this case may also pose problems.

\bigbreak\bigskip\bigskip\centerline{{\bf Acknowledgements}}\nobreak
Part of this work was done during a visit by one of us (TJ) to
the Aspen Center for Physics.
IJ and KLR were supported by PPARC via an Advanced Fellowship and
a Research Studentship respectively.   We also thank Hugh Osborn, who drew
our attention to Ref.~\ross .

\appendix{A}{The beta--functions and mass matrices}

In this appendix we make a few comments about the  $\beta$--functions
and sparticle mass matrices for the SSM, as defined in Eq.~\ssme .

The one--loop $\beta$--functions and the mass matrices appear
in many papers, and the two--loop $\beta$-functions are readily deduced
from the results of Ref.~\mv\ (or somewhat less readily from those
of Ref~\jja\ and~\yamada). These $\beta$--functions are calculated
in a ``hybrid'' regularisation scheme, intermediate, in a sense,
between dimensional regularisation and dimensional reduction. The
{\it raison d'\^etre\/} of the scheme is to remove annoying
dependence on $\epsilon$--scalar masses. The nature of the scheme
must be taken into account in the calculation of threshold corrections,
as explained in Ref~\jjmvy.

Although, as stated above, the one--loop
results have been often reproduced, we feel it worthwhile
emphasising the following point.
There are various possible conventions
for signs,  in particular of  $\mu_s$ and $M$,
and it is important, of course,
that the choices made in the $\beta$--functions are  consistent with
those made in the mass matrices.
We have verified all ``sensitive'' sign choices
by means of the identity
\eqn\zza{
\hbox{ STr }M^4 = 32\pi^2\left[\sum_{\la} \beta_{\la}^{(1)}.
{\pa\over{\pa\la}} -
\sum_{i=1,2}\ga_{H_i}^{(1)} . {\pa\over{\pa v_i}}\right] V_0 (v_1, v_2)}

where $V_0$ is the effective potential in the tree approximation,
i.e.\
\eqn\zzb{
 V_0 = \half (m_1^2 + \mu_s^2)v_1^2 + \half (m_2^2 + \mu_s^2)v_2^2
 - m_3^2 v_1 v_2+ {1\over 32}(g^2 + g'{}^2)(v_1^2 - v_2^2)^2.}

Eq.~\zza\ follows from the renormalisation group equation for the
effective potential. The set $\{\la\}$ consists of
$\{m_1^2, m_2^2, m_3^2, \mu_s, g, g'\}$.
The two anomalous dimensions $\ga_{H_i}^{(1)}$
are the one--loop anomalous dimensions for the background scalar
fields $H_{1,2}$ in the (quantum field) Landau gauge, given in a general
covariant gauge by
\eqn\zzc{\eqalign{
\lf\ga^{(1)}_{H_1} &= \la^2_{\tau} + 3 \la^2_b
- {1 \over 4} (1 + \alpha )(3g^2 + g' {}^2)\cr
\lf\ga^{(1)}_{H_2} &=3 \la^2_t - {1 \over 4} (1 + \alpha )(3g^2 + g' {}^2).\cr
}}
As usual $\alpha = 0$ is the Landau gauge. Note that for $\alpha = 1$
(the Feynman gauge), $\ga_{H_{1,2}}^{(1)}$ are identical to the corresponding
anomalous dimensions for the chiral superfields, in a supersymmetric gauge.
For completeness, we note that the corresponding anomalous dimensions
for the quantum scalar fields, $\gatilde_{H_{1,2}}^{(1)}$, are given by
\eqn\zzd{\eqalign{
\lf\gatilde^{(1)}_{H_1} &= \la^2_{\tau} + 3 \la^2_b
- {1 \over 4} (1 - \alpha )(3g^2 + g' {}^2)\cr
\lf\gatilde^{(1)}_{H_2} &=3 \la^2_t
- {1 \over 4} (1 - \alpha )(3g^2 + g' {}^2),\cr
}}
differing only in the sign of the gauge parameter term.
\listrefs
\bye